\documentclass[twocolumn]{aastex61}
\usepackage{natbib}

\newcommand{\ammonia}{$\mbox{NH}_3$}
\newcommand{\kms}{$\mbox{km~s}^{-1}$}
\newcommand{\ionhy}{H{\sc ii} }
\newcommand{\msol}{\mbox{M\hbox{$_{\odot}$}}}

\shorttitle{84\,GHz class I methanol maser emission towards NGC\,253}
\shortauthors{McCarthy et al.}

\begin{document}
	
	\title{Detection of 84\,GHz class I methanol maser emission towards NGC\,253}
	
	\correspondingauthor{Tiege McCarthy}
	\email{tiegem@utas.edu.au}
	
	\author[0000-0001-9525-7981]{Tiege P. McCarthy}
	\affil{School of Natural Sciences, University of Tasmania, Hobart, Tasmania 7001, Australia}
	\affil{Australia Telescope National Facility, CSIRO, PO Box 76, Epping, NSW 1710, Australia}
	
	\author[0000-0002-1363-5457]{Simon P. Ellingsen}
	\affil{School of Natural Sciences, University of Tasmania, Hobart, Tasmania 7001, Australia}
	
	\author[0000-0002-4047-0002]{Shari L. Breen}
	\affiliation{Sydney Institute for Astronomy (SIfA), School of Physics, University of Sydney, NSW 2006, Australia}
	
	\author[0000-0002-4047-0002]{Maxim A. Voronkov}
    \affiliation{Australia Telescope National Facility, CSIRO, PO Box 76, Epping, NSW 1710, Australia}	
    
	\author[0000-0002-5435-925X]{Xi Chen}
	\affiliation{Center for Astrophysics, GuangZhou University, Guangzhou 510006, China}
	\affiliation{Shanghai Astronomical Observatory, Chinese Academy of Sciences, Shanghai 200030, China}

\begin{abstract}
\noindent 

We have investigated the central region of NGC~253 for the presence of 84.5~GHz ($5_{-1}\rightarrow4_0$E) methanol emission using the Australia Telescope Compact Array. We present the second detection of 84.5~GHz class~I methanol maser emission outside the Milky Way. This maser emission is offset from dynamical centre of NGC~253, in a region with previously detected emission from class~I maser transitions (36.2~GHz $4_{-1}\rightarrow3_0$E and 44.1~GHz $7_{0}\rightarrow6_1$A$^{+}$ methanol lines) . The emission features a narrow linewidth ($\sim$12~\kms) with a luminosity approximately 5 orders of magnitude higher than typical Galactic sources. We determine an integrated line intensity ratio of $1.2\pm0.4$ between the 36.2~GHz and 84.5~GHz class~I methanol maser emission, which is similar to the ratio observed towards Galactic sources. The three methanol maser transitions observed toward NGC~253 each show a different distribution, suggesting differing physical conditions between the maser sites and that observations of additional class~I methanol transitions will facilitate investigations of the maser pumping regime.

\end{abstract}

\section{Introduction} \label{sec:intro}

Methanol maser emission provides a powerful tool for investigating star-formation regions within the Milky Way. Methanol maser transitions are empirically divided into two classes, based on the pumping mechanism responsible for population inversion \citep{Batrla+87, Menten91a}. Class~I methanol masers are pumped via collisional interactions, whereas the class~II transitions are radiatively pumped. Both classes of methanol masers are commonly observed towards Galactic star-formation regions, with over 1200 unique sources reported \citep[e.g.][]{Ellingsen+05,Caswell+10,Caswell+11, Voronkov+14, Breen+15, Green+10, Green+12a, Green+17}. The radiatively pumped class~II masers are observed to be exclusively associated with young-stellar objects (YSOs) in high-mass star-formation regions \citep{Breen+13b}, while class~I masers are observed further from the excitation sources in shocked gas from molecular outflows or expanding \ionhy regions \citep{Kurtz+04, Cyganowski+09, Cyganowski+12, Voronkov+10a, Voronkov+14}.

Extragalactic methanol masers have only been reported in a handful of sources across both classes. Class~II emission has been observed (in the 6.7 and 12.2~GHz transitions) towards two local group galaxies, the Large Magellanic Cloud (LMC) and M31 \citep{Green+08,Ellingsen+10,Sjouwerman+10} and the nearby starburst NGC~4945 (Ellingsen et al., submitted). There are 6 sources outside of the Milky Way with detections of class~I methanol masers, NGC~253, Arp~220, NGC~4945, IC~342 and NGC~6946 at 36.2~GHz \citep{Ellingsen+14, Chen+15, McCarthy+17, Gorski+18}, NGC253 at 44.1~GHz \citep{Ellingsen+17b}, and NGC~1068 at 84.5~GHz \citep{Wang+14}. The extragalactic class~II methanol maser emission towards both M31 and the LMC appears similar to the class~II masers observed in our Galaxy, indicating they are simply bright examples of Galactic-style class~II methanol masers. In contrast, the extragalactic class~I emission appears to show significant departure from the properties of their Galactic counterparts, with the current hypothesis being they occur due to large-scale low velocity shocks driven by dynamical processes such as molecular in-flow along galactic bars \citep{Ellingsen+14, Ellingsen+17b, Gorski+18}. These molecular in-flows may be related to the starburst/star-formation of the host galaxy, presenting the possibility that the methanol maser emission scales with star-formation \citep{Chen+16, Ellingsen+17b,Gorski+18}.

The plethora of transitions and unique pumping mechanisms between the two discrete classes of methanol masers makes them useful tools for studying various astrophysical phenomena. Class II methanol masers have been shown to be only associated with high-mass star formation regions and hence can be used as a direct signpost for the location of the newly formed high-mass star \citep{Breen+13b}.  They have also been used to investigate the kinematics of the gas in these regions at high resolution \citep[e.g. ][]{Goddi+11}.  The class~I transitions trace outflows and shock fronts in the extended regions of these stellar nurseries \citep{Kurtz+04,Cyganowski+12,Voronkov+14,McCarthy+18}. The various transitions of methanol can also be used to investigate potential changes in the fundamental physical constants, such as the proton-to-electron mass ratio \citep{Levshakov+11}. Observations and comparison of multiple methanol maser transitions from distant galaxies will allow investigation into any variations in this mass ratio across cosmic time (see Section \ref{sec:pe_mass_ratio} for further discussion).

NGC~253 is a barred-spiral starburst galaxy in the nearby Sculptor group. Due to its proximity \citep[3.4~Mpc; ][]{Dalcanton+09} and starburst nucleus, it has been extensively studied across a broad range of wavelengths, and displays emission from numerous molecular species \citep{Martin+06b, Meier+15, Ellingsen+17b}. A star-formation rate of $\sim1.7~\msol~$yr$^{-1}$ has been determined for the central starburst region of NGC~253 \citep{Bendo+15}. A variety of extragalactic maser species have been observed towards NGC~253, with OH, H$_2$O and \ammonia\ masers having been detected towards the nucleus \citep{Turner+85,Henkel+04, Hofner+06, Ott+05, Gorski+17} along with 36.2~and~44.1~GHz methanol and HC$_3$N masers distributed across the central molecular region \citep{Ellingsen+17,Ellingsen+17b}. 

\citet{Huttemeister+97} used the IRAM 30 m telescope to map the central region of NGC253 in a number of methanol and formaldehyde transitions in the 3 mm band. They detected 3 different methanol transitions towards this region, the $5_{-1}\rightarrow4_0$E (84.5~GHz line), $2_k \rightarrow 1_k$ (96.7~GHz) and $3_k \rightarrow 2_k$ (145.1~GHz) lines. Although \citeauthor{Huttemeister+97} found it difficult to reconcile the properties of the 84.5~GHz line with the other observed thermal transitions, they did not consider that it may be the result of maser emission (despite it being a class~I transition in Galactic star-formation regions). Given the recent detection of maser emission from the related 36.2~GHz methanol transition, higher resolution observations of this 84.5~GHz transition were warranted in order to both determine its nature and and more accurately determine its location relative to other methanol maser lines.


We have conducted 3~mm observations towards NGC~253 in order to investigate the 84.5~GHz methanol maser line. This line is directly related to the 36.2~GHz class~I masing transition (in the same transition family), which has been previously detected towards this source \citep{Ellingsen+14,Ellingsen+17b}. In Galactic sources these two masers are frequently observed together towards the same sources \citep[][Breen et al. in preparation]{Haschick+89,Kalenskii+01,Fish+11,Voronkov+06,Voronkov+14}. It has been speculated on multiple occasions that that the 84.5 GHz line may be present in sources with observed extragalactic 36.2~GHz methanol maser emission. In addition to the 84.5~GHz methanol line, we also simultaneously observed the 87.9~GHz HNCO line, which is a tracer of low-velocity shocks \citep{Meier&05}. 


\section{Observations} \label{sec:obs}


The 3\,mm observations were made using the Australia Telescope Compact Array (ATCA) during Director's time on 2018 July 27 and 30 (project code C3167). The observations used a hybrid array configuration (H75) with maximum and minimum baselines of 89~m and 31~m respectively. Antenna 6 is not fitted with a 3 mm receiver, therefore, it was excluded from the observations. The Compact Array Broadband Backend \citep[CABB ;][]{Wilson+11} was configured in CFB 64M-32k mode. This mode consists of two 2 GHz IF bands, with $32 \times 64$~MHz channels, and up to 16 of these 64~MHz channels can be configured as zoom bands with $2048 \times 31.2$~kHz channels. The primary science target of these observations was the 84.5 GHz methanol maser transition, for which we adopted a rest frequency of 84\,521.206$\pm0.001$~MHz \citep{Zuckerman+72, Xu+97}. The $31.2$~kHz spectral resolution corresponds to a velocity resolution of 0.11~\kms\ at 84.5~GHz. Along with the $5_{-1}\rightarrow4_0$E methanol line, we also observed the 87.9~GHz HNCO $4_0\rightarrow3_0$ transition \citep[rest frequency of 87\,925.238~MHz;][]{Turner91}.

In addition to the primary 3\,mm observations, we also observed NGC~253 at 7\,mm on 2018 August 1 during Director's time in order to monitor emission from the 36.2~GHz $4_{-1}\rightarrow3_0$E methanol maser line. The observing configuration was not optimal for high-resolution imaging, as the compact configuration (H75) does not provide high enough angular resolution at 7\,mm (compared to already existing data on the 36.2~GHz line). The CABB configuration for these observations was identical to the 3\,mm observations described previously, with an adopted rest frequency of $36\,169.238\pm0.001$~MHz \citep{Voronkov+14} and a spectral resolution of 0.26~\kms.

PKS~B1921-293 was used as the bandpass calibrator for both sets of observations, flux-density was calibrated with respect to Uranus (for 3\,mm observations) and PKS B1934-648 (for 7\,mm), while the phase calibration source was PKS~B0116-219. The observing strategy interleaved 600 seconds on the target source with 100 seconds on the phase calibrator. The data were corrected for atmospheric opacity and the absolute flux density calibration is estimated to be accurate to better than 30\%. System temperatures were tracked using measurements of a paddle at ambient temperature for the 3\,mm observations (relying on assumptions of unchanged atmospheric conditions between paddle scans), whereas a noise diode and model for atmospheric opacity (in {\sc miriad}) was used at 7\,mm. The rms pointing error across all telescopes utilised was approximately 3.8 arcseconds. The total on-source time for NGC~253 is 307 and 247 minutes, for 3\,mm (both days) and 7\,mm observations respectively. 

{\sc miriad} was used for data reduction, following standard techniques for the reduction of ATCA 3\,mm and 7\,mm spectral line observations. Self-calibration was performed on the data using the continuum emission from the line-free channels toward the central region of NGC~253 (both phase and amplitude). The continuum emission was subtracted from the self-calibrated uv-data with the uvlin task, which estimates the intensity on each baseline from the line-free spectral channels. This enables us to isolate any spectral line emission from continuum emission. The systemic velocity of NGC~253 is 243~\kms\ \citep[Barycentric;][]{Koribalski+04} and our 3\,mm image cube covered a range of 0 to 500~\kms. The spectral line cube for the 36.2~GHz line covers a velocity range of 120 to 470~\kms. However, this narrower range was still sufficient to cover the whole velocity range over which 36.2~GHz methanol emission had previously been observed \citep{Ellingsen+17b}. The spectral line data was resampled and imaged with a channel width of 6~\kms\ for both of the frequency setups. The {\sc miriad} task imfit was use to determine positions and peak flux densities of the spectral line emission. This task reports the peak value and location of a two-dimensional Gaussian fit for the emission in a given velocity plane within the spectral line cube. Because it reports the parameters of the Gaussian fit, minor differences may be present between the reported flux-density values, and those apparent from the extracted spectra. A Brigg's visibility weighting robustness parameter of 1 was used when generating the spectral line cubes with {\sc miriad}. This resulted in a synthesised beam for our 3\,mm observations of approximately $6\farcs4 \times 4\farcs9$ with a position angle of $-77.1^\circ$, and $17\farcs4 \times 11\farcs5$ with position angle of $-69.3^\circ$ for our 7\,mm observations.

\begin{table*}
	\begin{center}
		\caption{Properties of methanol and HNCO emission from our observations towards NGC~253. The locations and peak flux densities given are those of the peak emission components in each transition as extracted using the imfit {\sc miriad} task on the spectral line cubes. Positional uncertainties (from fitting to the emission components) for the 3\,mm emission is accurate to approximately 0.5 arcsecond, and 7\,mm emission to approximately 1 arcseconds. All tabulated velocities are with respect to the Barycenter.}
		\begin{tabular}{llcllccccc} \hline
			\multicolumn{1}{c}{\bf} & \multicolumn{1}{c}{\bf Rest}& \multicolumn{1}{c}{\bf Location} & \multicolumn{1}{c}{\bf R.A.}  & \multicolumn{1}{c}{\bf Dec.} & \multicolumn{1}{c}{\bf Peak} & \multicolumn{1}{c}{\bf Integrated} & \multicolumn{1}{c}{\bf Peak} & \multicolumn{1}{c}{\bf Velocity} & \multicolumn{1}{c}{\bf RMS} \\
			\multicolumn{1}{c}{\bf} & \multicolumn{1}{c}{\bf Frequency} & \multicolumn{1}{c}{\bf Reference} & \multicolumn{1}{c}{(J2000)}  & \multicolumn{1}{c}{(J2000)} & \multicolumn{1}{c}{\bf Flux Density} & \multicolumn{1}{c}{\bf Flux Density} & \multicolumn{1}{c}{\bf Velocity} & \multicolumn{1}{c}{\bf Range} & \multicolumn{1}{c}{\bf Noise} \\
			& \multicolumn{1}{c}{(GHz)} & &  \multicolumn{1}{c}{\bf $h$~~~$m$~~~$s$}& \multicolumn{1}{c}{\bf $^\circ$~~~$\prime$~~~$\prime\prime$} & (mJy) & (mJy\,km\,s$^{-1}$) & (km\,s$^{-1}$) & (km\,s$^{-1}$) & (mJy) \\   \hline
			CH$_3$OH & 84.521206$^{[1]}$ & B & 00 47 33.65 & $-$25 17 12.6 & 30.2 & $693\pm50$  & 187 & 134 -- 216 & 2.6 \\ 
			& 36.169238$^{[3]}$ & NE & 00 47 33.90 & $-$25 17 11.6 & 30.0 & $961\pm36$  & 211 & 163 -- 235 & 0.5 \\
			& & SW & 00 47 32.00 & $-$25 17 28.0 & 27.0 & $1372\pm46$  & 304 & 262 -- 355 & 0.5 \\ 
			HNCO & 87.925238$^{[2]}$ & A & 00 47 33.94 & $-$25 17 11.0 & 53.9 & $953\pm76$ & 205 & 187 -- 223 & 5.1 \\
			& & B & 00 47 33.49 & $-$25 17 13.8 & 41.3 & $458\pm42$ & 181 & 163 -- 187 & 4.0 \\
			& & C & 00 47 32.79 & $-$25 17 21.9 & 23.7 & $276\pm50$ & 289 & 277 -- 301 & 4.3 \\
			& & D & 00 47 32.29 & $-$25 17 20.5 & 67.5 & $1502\pm88$ & 337 & 319 -- 361  & 4.6 \\
			& & E & 00 47 31.96 & $-$25 17 27.2 & 50.3 & $1614\pm125$ & 301 & 271 -- 331 & 3.7 \\ \hline
		\end{tabular} \label{tab:emission}		
	\end{center}
	\begin{flushleft}
		Note: $^{[1]}$\citet{Xu+97}, $^{[2]}$\citet{Turner91} and $^{[3]}$\citet{Voronkov+14} \\
	\end{flushleft}	
\end{table*}

\begin{table*}
	\begin{center}
		\caption{Properties of continuum emission from our observations towards NGC~253. Positional uncertainties (from fitting to the emission components) for the 3\,mm emission is accurate to approximately 0.5 arcsecond, and 7\,mm emission to approximately 1 arcseconds.}
		\begin{tabular}{lcccccc} \hline
			\multicolumn{1}{c}{\bf} & \multicolumn{1}{c}{\bf Frequency} &\multicolumn{1}{c}{\bf R.A.}  & \multicolumn{1}{c}{\bf Dec.} & \multicolumn{2}{c}{\bf Flux Density}  & \multicolumn{1}{c}{\bf RMS} \\
			\multicolumn{1}{c}{\bf}  & \multicolumn{1}{c}{\bf Range} & \multicolumn{1}{c}{(J2000)}  & \multicolumn{1}{c}{(J2000)} & \multicolumn{1}{c}{\bf Peak} & \multicolumn{1}{c}{\bf Integrated} & \multicolumn{1}{c}{\bf Noise} \\
			&\multicolumn{1}{c}{(GHz)} &\multicolumn{1}{c}{\bf $h$~~~$m$~~~$s$}& \multicolumn{1}{c}{\bf $^\circ$~~~$\prime$~~~$\prime\prime$} & (mJy) & (mJy)  & (mJy) \\   \hline
			3\,mm continuum & $84.376 - 84.536$  & 00 47 33.06 & $-$25 17 18.4 &  149 &$217\pm40$  & 0.6 \\
			7\,mm continuum & $36.064 - 36.160$  & 00 47 33.06 & $-$25 17 18.4 &  369 & $380\pm24$  &  0.1 \\   \hline
		\end{tabular} \label{tab:continuum}		
	\end{center}
\end{table*}

\begin{figure*}
	\begin{minipage}[h]{0.24\linewidth}
		\centering
		\includegraphics[scale=0.48]{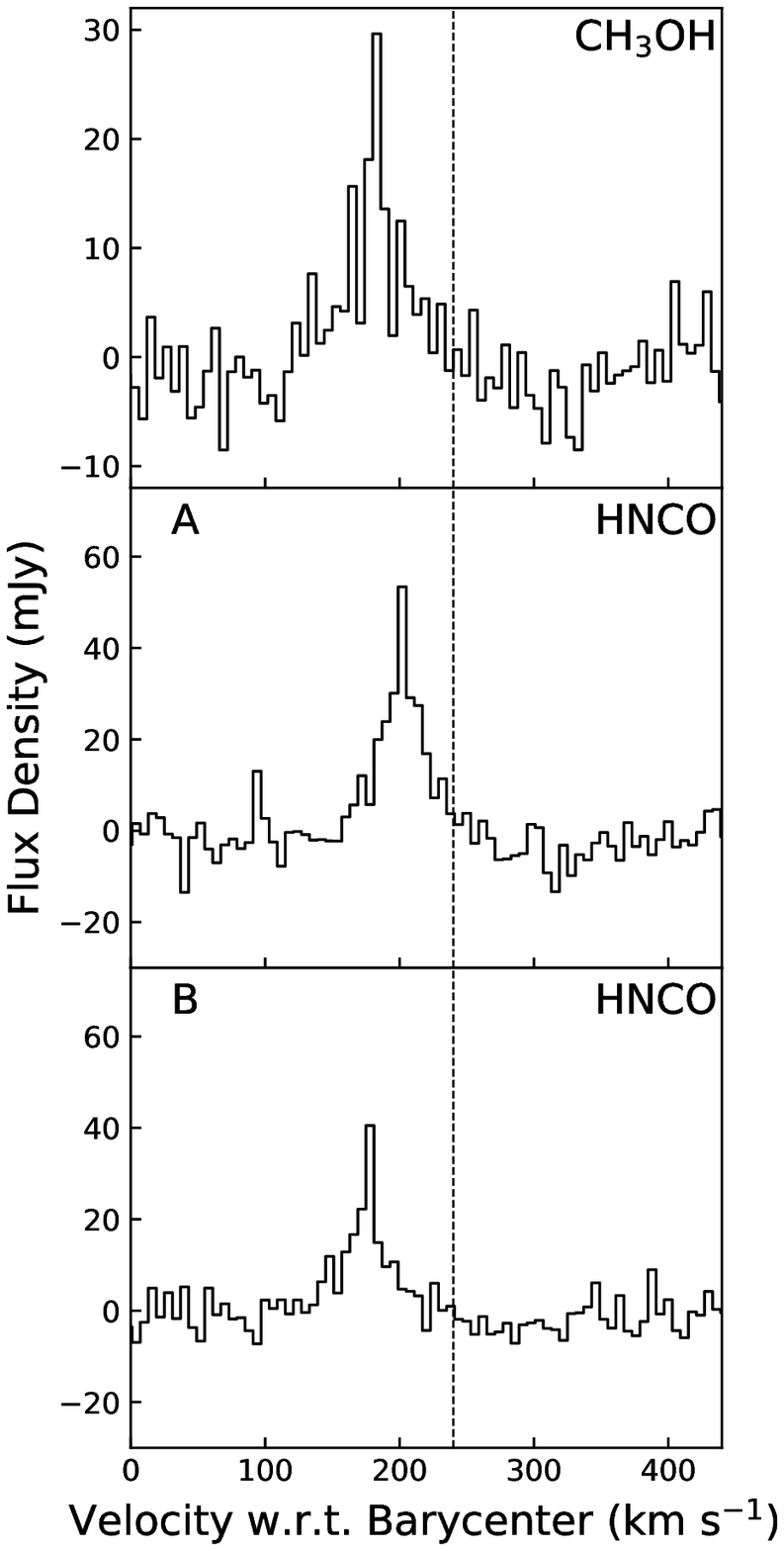}
	\end{minipage}
	\begin{minipage}[h]{0.48\linewidth}
		\centering
		\includegraphics[scale=0.40]{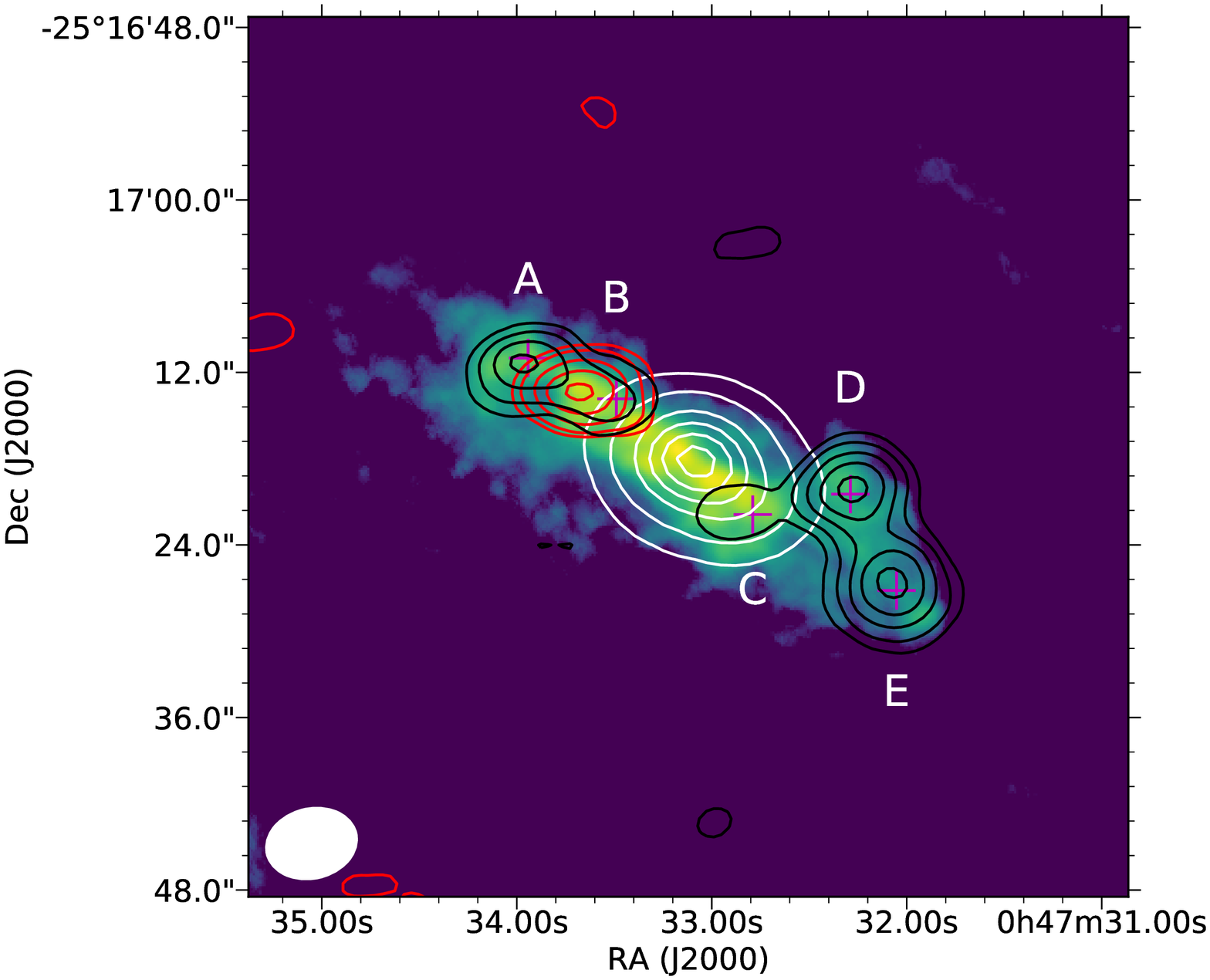}
	\end{minipage}
	\begin{minipage}[h]{0.25\linewidth}
		\centering
		\includegraphics[scale=0.48]{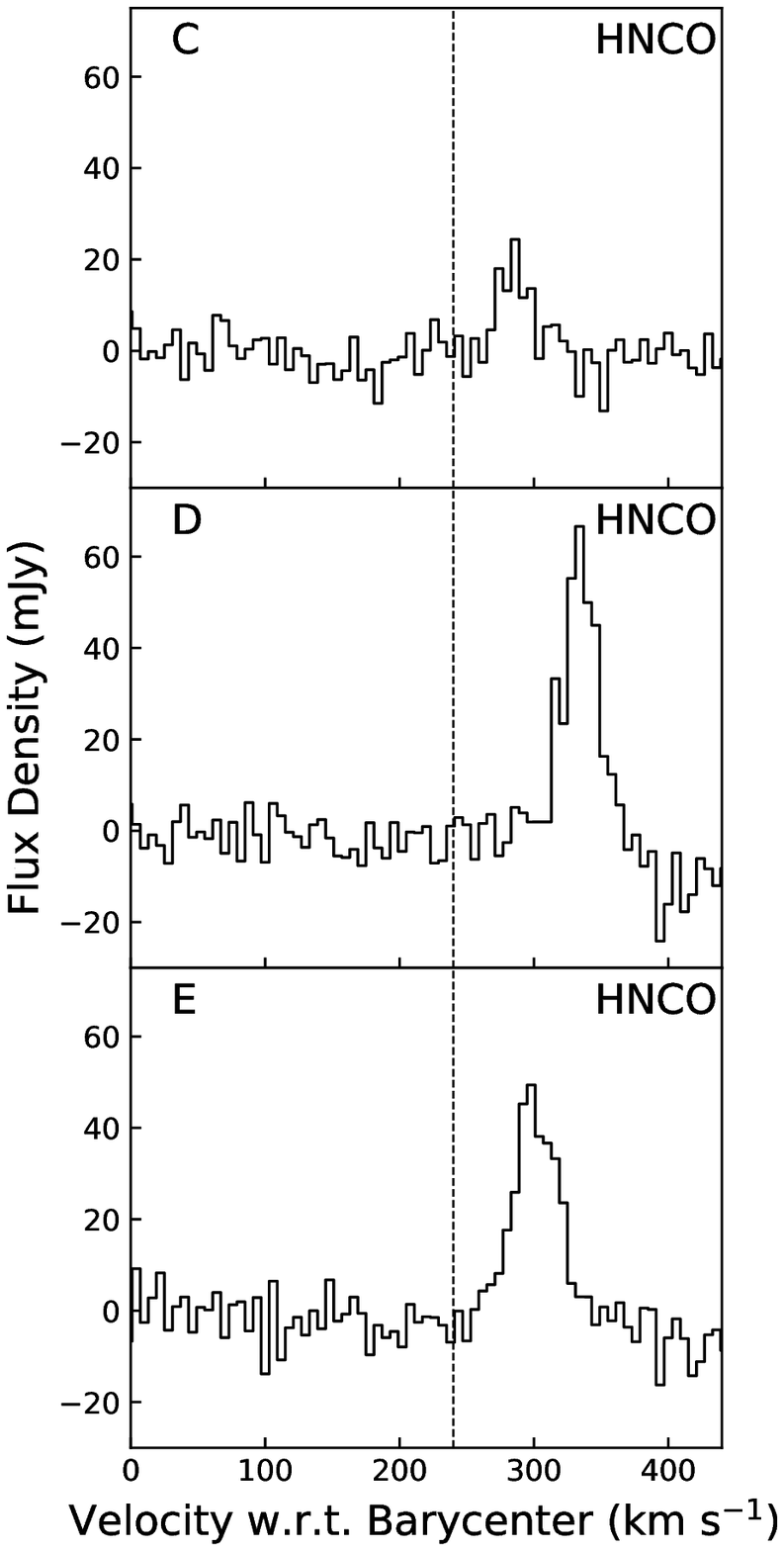}
	\end{minipage}
	\caption{Middle: Integrated 84.5\,GHz methanol emission (red contours 10\%, 30\%, 50\%, 70\%, and 90\% of the 573 mJy\,\kms ~beam$^{-1}$ peak), integrated 87.9~GHz HNCO emission (black contours 15\%, 30\%, 50\%, 70\%, and 90\% of the 1.9 Jy\,\kms ~beam$^{-1}$) and the 3mm continuum emission (white contours 2\%, 10\%, 30\%, 50\%, 70\%, and 90\% of the 149 mJy~beam$^{-1}$ peak) with background image of integrated CO $J = 2 \rightarrow 1$ emission from \citet{Sakamoto+11} on a logarithmic scale. Magenta plus signs indicate the peak components of HNCO emission and have been labelled A through E moving from high right ascension to low. Left: Spectra for blue shifted emission from the region of 84.5~GHz methanol emission as well as HNCO regions A and B. Right: Spectra for red shifted emission from the HNCO regions C, D and E. All spectra extracted from the relevant spectral line cubes (imaged at 6\,km\,s$^{-1}$). The vertical dashed line indicates the systemic velocity of NGC\,253 \citep[243~\kms;][]{Koribalski+04}.}
	\label{fig:plot&spec}
\end{figure*}

\section{Results} \label{sec:results}

We have detected 84.5~GHz ($5_{-1}\rightarrow4_0$E) and 36.2~GHz ($4_{-1}\rightarrow3_0$E) methanol emission and 87.9~GHz ($4_{0}\rightarrow3_0$) HNCO emission along with 3\,mm and 7\,mm continuum emission towards NGC~253. Details of the detected spectral line and continuum, emission are tabulated in Table \ref{tab:emission} and Table \ref{tab:continuum} respectively.

The 3~mm continuum emission is approximately centred on the nucleus of NGC~253. Compared to the 7~mm continuum peak observed here, and by \citet{Ellingsen+17b}, we see the 3~mm peak offset to the west by approximately 0.5 arcseconds. Given that this offset is on the order of the typical astrometric uncertainty achieved with ATCA, we cannot state with certainty that the offset between the 3\,mm and 7\,mm continuum sources is the result of a real spatial separation.

\begin{figure}
	\includegraphics[width=\linewidth]{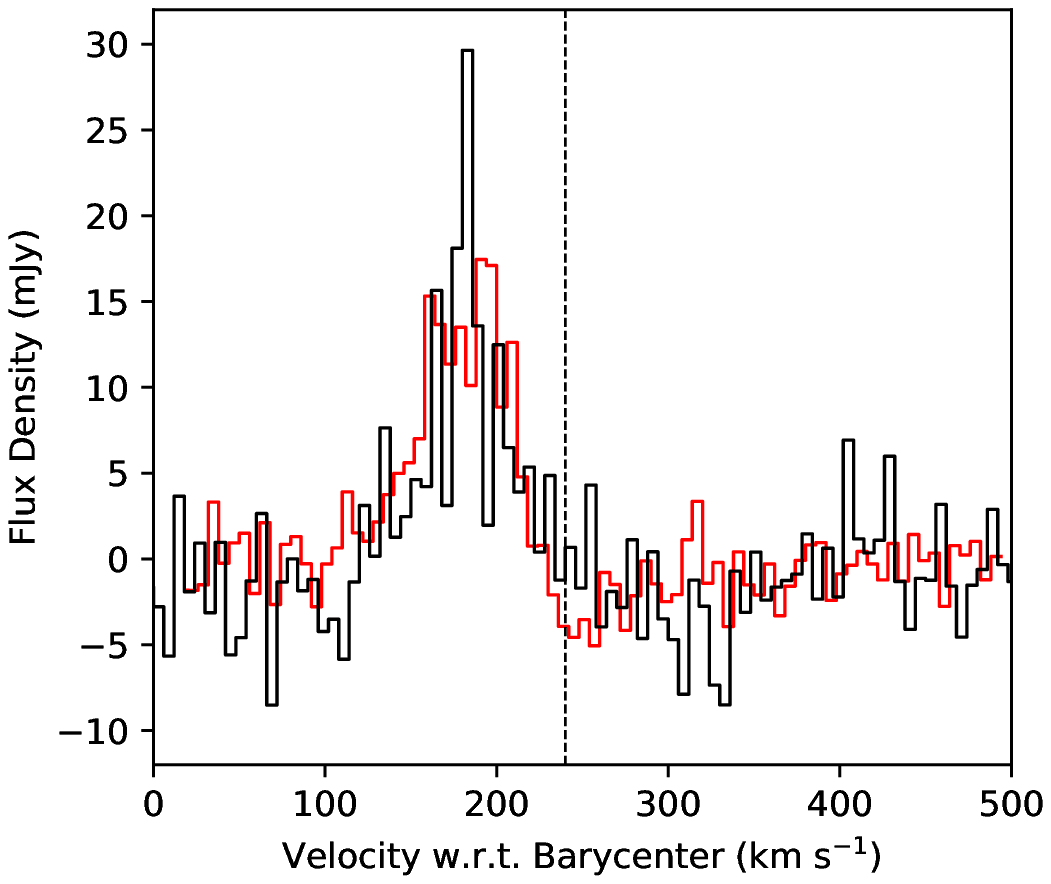}
	\caption{84.5~GHz maser emission (black) with 36.2~GHz methanol maser emission (red) from the same region (MM4) superimposed. The 36.2~GHz emission is from \citet{Ellingsen+17b}, however we have restored it with the same $6\farcs4 \times 4\farcs9$ beam from our 84.5~GHz observations.}
	\label{fig:84_36_overlay}
\end{figure}

The 84.5~GHz methanol emission is located north-east of the continuum peak (see Figure \ref{fig:plot&spec}). This location corresponds to the 36.2~GHz methanol maser region MM4/MM5, as defined by \citet{Ellingsen+17b}. The 84.5~GHz methanol is not only observed from the same location, it also covers a very similar velocity range to the 36.2~GHz methanol masers at this location (see Figure \ref{fig:84_36_overlay}). The spectrum appears to result from a single component at $\sim187$~\kms\, which is in good agreement with the spectral profile presented in \citet{Huttemeister+97}, although the peak flux density obtained for the ATCA data is a factor of 6 lower than that reported by \citeauthor{Huttemeister+97} (assuming a conversion factor of 6.0~Jy/K for mid 90s IRAM data).

Multiple components of 87.9~GHz HNCO emission are observed, all offset from the 3~mm continuum peak emission and approximately aligned with the plane of the disk (see Figure \ref{fig:plot&spec}). The locations of the components are consistent with the higher-resolution ALMA observations of the same transition by \citet{Meier+15}, and the lower frequency transition by \citet{Ellingsen+17b}, with a component of the HNCO emission observed at the location of the 84.5~GHz methanol emission. We have labelled the 5 components of HNCO as A through E (see Figure \ref{fig:plot&spec}).

\begin{figure}
	\includegraphics[width=\linewidth]{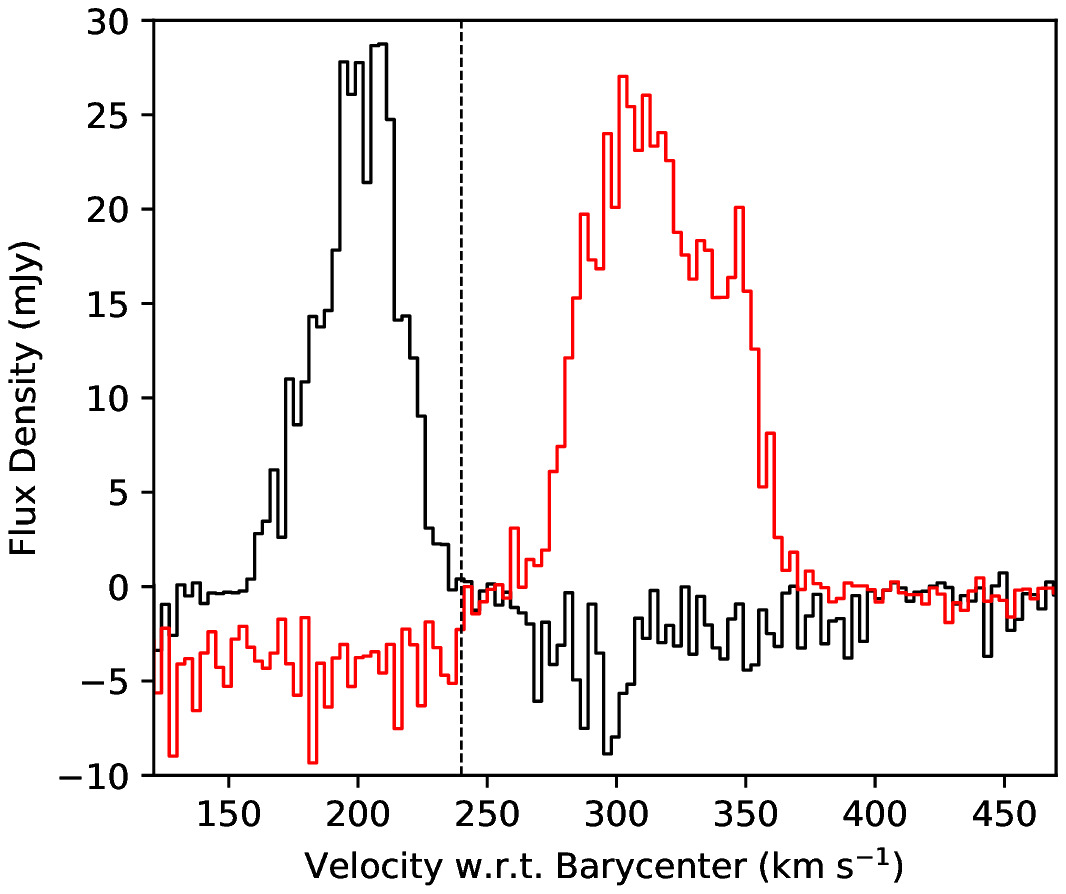}
	\caption{36.2~GHz methanol maser spectrum towards NGC~253 from our ATCA H75 observations. Both spectra were taken from the self-calibrated spectral line cube with a resolution of 3~\kms. Emission towards the north-east and south-west regions are represented by the black and red spectra respectively. }
	\label{fig:36}
\end{figure}

The 36.2~GHz methanol maser transition was detected towards NGC~253 both north-east and south-west of the continuum emission at approximately the same locations described in Figure 1 of \citet{Ellingsen+14}. We have not included an image of this line due to the coarse angular resolution of our 36.2~GHz observations (in comparison to existing 36.2~GHz observations towards NGC~253), however, the spectra at the two locations are shown in  Figure \ref{fig:36}. These observations have been included here as they are contemporaneous with the observations at 84.5~GHz and show that the emission from the 36.2~GHz transition is largely unchanged from previous observations \citep{Ellingsen+14,Ellingsen+17b}. 

Further comparison between 84.5~ and 36.2~GHz emission presented in the following sections uses the intermediate resolution (original synthesised beam of $3\farcs4 \times 0\farcs9$ with a position angle of $7.8^\circ$) data presented in \citet{Ellingsen+17b} restored with the same synthesised beam size as the 84.5~GHz observations ($6\farcs4 \times 4\farcs9$ with a position angle of $-77.1^\circ$). This higher resolution data, convolved with a restoring beam with the same dimensions as the 84.5 GHz data allows for a more accurate comparison of the two transitions.  However, as the two transitions were observed at different epochs, some uncertainty remains.

\section{Discussion} \label{sec:disc}

\subsection{The nature of the 3\,mm methanol: masing or thermal}

Class~I methanol maser emission has previously been detected in the 36.2 and 44.1~GHz methanol class~I transitions towards NGC~253 \citep{Ellingsen+14,Ellingsen+17b}. The 84.5~GHz methanol transition we observe here is from the same transition family as the 36.2~GHz emission. In Galactic sources, these two lines are frequently observed together as masers (Breen et al. in preparation). This, combined with the fact that 84.5~GHz maser emission has been previously detected in an external galaxy \citep{Wang+14}, the single dish detection of 84~GHz methanol by \citet{Huttemeister+97}, and our tentative detection from the Mopra 22\,m telescope, prompted our search for 84.5~GHz emission towards NGC~253. Our angular resolution is a factor of 5 times higher than the \citet{Huttemeister+97} observations and allows us to better constrain the nature of the emission. Like the previous detections of methanol maser emission, we need to adequately justify whether the emission we are observing here is due to thermal or maser processes.

The offset region from which we observe the 84.5~GHz methanol emission is also the location of previously reported 36.2~ and 44.1~GHz masers \citep[MM4; ][]{Ellingsen+17b}. The 36.2~ and 84.5~GHz methanol emissions extends over the same velocity range, with the peak flux-density occurring at approximately the same velocity (see Figure \ref{fig:84_36_overlay}). The 44.1~GHz maser component is also at approximately at the same velocity as these other two transitions, however, it covers a much more narrow velocity range \citep{Ellingsen+17b}. \citet{Ellingsen+17b} present comprehensive arguments for both why the 36.2~ and 44.1~GHz emission in these locations are masers, and that they can not be the result of large-scale emission from numerous Galactic-style masers. In addition, high-angular resolution JVLA observations (0.1 arcseconds) of this region have shown that the brightness temperature of some of the 36 GHz emission at this location, exceeds 1000 K, demonstrating conclusively that it is a maser \citep[location B in][]{Chen+18}. The similarity between the emission from all three of these transitions indicates the 84.5~GHz emission is likely the result of a maser process. 

The median integrated flux density of Galactic 84.5~GHz class~I maser sources is 29.4~Jy\,\kms\ (Breen et al. in preparation). This corresponds to an isotropic luminosity of 735~Jy\,\kms\,kpc$^2$ ($5.13 \times 10^{-6}$\,\mbox{L\hbox{$_\odot$}}) at a distance of 5~kpc. Assuming a distance of 3.4~Mpc to NGC~253 \citep{Dalcanton+09}, we determine an isotropic luminosity for this region of $6.6 \times 10^6$~Jy\,\kms\,kpc$^2$ ($0.046$\,\mbox{L\hbox{$_\odot$}}), almost nine thousand times more luminous than the typical Galactic source. This implies that the luminosity of 84.5~GHz emission from this single region towards NGC~253, exceeds the combined luminosity of all Galactic 84.5~GHz emission.

The peak emission component at 84.5~GHz has a FWHM of approximately 12~\kms. Linewidths of the thermal emission from this location are much broader than what we are observing from the 84.5~GHz methanol emission \citep{Huttemeister+97, Leroy+15}. This narrow linewidth and higher peak flux density, when compared to the 36.2~GHz maser emission, suggest that this emission comes from a region with a smaller total volume and higher gain. Typically maser processes are required in order to observe narrow linewidths from transitions with relatively high upper-state energy levels (such as the $5_{-1}\rightarrow4_0$E transition considered here). In addition comparison of the integrated line ratios between the 84.5~GHz methanol emission, the two other class~I transitions, and 48.4~GHz thermal methanol (Ellingsen et al. in preparation) from the same region are inconsistent with a thermal process. \citet{Huttemeister+97} found inconsistencies in their calculations of environmental properties when considering emission from the $5_{-1}\rightarrow4_0$E line as thermal, this would be expected if some component of the emission was resulting from maser processes. From this point on, we will refer to this emission as a class~I methanol maser.

\subsection{84.5~GHz methanol and probing the proton-to-electron mass ratio} \label{sec:pe_mass_ratio}

\citet{Ellingsen+17b} outline the potential use of extragalactic class~I methanol masers as probes of the proton-to-electron mass ratio. The rest frequencies of methanol transitions are sensitive to the proton-to-electron mass ratio due to the hindered internal rotation of the OH radical in the methanol molecule \citep{Levshakov+11}. Variations in these fundamental constants can be revealed through comparison of rest-frequencies observed astronomically and those observed in the laboratory \citep{Kanekar11, Bagdonaite+13}. For this to be practical on cosmological scales, multiple methanol transitions need to be identified that are sufficiently luminous, co-spatial and have different dependencies on these physical constants \citep[see Section 4.4 in][ for more detailed discussion]{Ellingsen+17b}. \citeauthor{Ellingsen+17b} conclude that the 44.1~GHz methanol maser lines observed towards NGC~253 are not appropriate for this kind of comparison, due to the significant difference in luminosity and spectral profile when compared to the 36.2~GHz emission. From our observations the 84.5~GHz transition broadly appears to be more appropriate for this sort of comparison \citep[with sensitivity coefficients of $-3.5$ and $-9.6$ for the 85.6 and 36.2~GHz transitions respectively;][]{Levshakov+11}. However, when closely comparing the 84.5~ and 36.2~GHz emission from the same region in NGC~253, we see that the 36.2~GHz is broken into two separate components, neither at the exact location or velocity of the 84.5~GHz component we observe \citep{Ellingsen+17b}. Higher angular resolution investigation into the 84.5~GHz transition will allow us to both; more accurately determine its position relative to the 36.2~GHz methanol emission (for which we already have sub-arcsecond accurate positions) and potentially resolve the emission into multiple components (which may or may not overlap in velocity with the 36.2~GHz components). If extragalactic class~I maser luminosities scale with star formation rate, we will be able to sensitively probe the fundamental constants of physics across cosmic time.

\subsection{Comparison with other class~I emission} \label{sec:compare}

When considering the \citet{Ellingsen+17b} intermediate resolution observations of 36.2~GHz towards the same region as the 84.5~GHz maser, we see two 36.2~GHz peaks either side of the 84.5~GHz peak velocity. This is due to the 36.2~GHz emission from this region consisting of two distinct components, one on the south side and the other on the west side of the MM4 region. At the resolution of our observations we are not able to determine whether the 84.5~GHz emission is broken into two discrete components also, with the spectrum we are seeing representing contributions from the two components. Figure \ref{fig:84_36_overlay} shows a composite spectrum with emission from both the 36.2~ and 84.5~GHz transition. This spectrum has been taken at the location of the 84.5~GHz peak (see Table \ref{tab:emission}), using both our 84.5~GHz data, and the intermediate resolution 36.2~GHz spectral line cube from \citet{Ellingsen+17b} restored with the same synthesised beam size (see Section \ref{sec:results} for further details).

The integrated flux-density of this re-convolved 36.2~GHz emission is 798 mJy~\kms, resulting in a integrated intensity ratio between the 36.2/84.5~GHz lines of $1.2\pm0.4$. This ratio is similar to the median 36.2/84.5 GHz ratio observed towards Galactic sources of 1.4 (Breen et al. in preparation). In order to determine the 36.2/84.5~GHz line ratio across the remaining methanol maser regions, we determine a generous upper limit for the integrated flux density of undetected 84.5~GHz emission of 326 mJy~\kms\ (for emission with 5$\sigma$ peak and FWHM of 10~\kms). We compare this upper limit 84.5~GHz integrated flux against 36.2~GHz integrated flux densities from our re-convolved cube at the locations of the 36.2~GHz components \citep[see Table 2 of][]{Ellingsen+17b}. This results in lower limit line ratios of 3.5, 4.2, 2.9, 1.7 and 1.5 for regions MM1, MM2, MM3, MM6 and MM7 respectively. This indicates that in all of these regions, the intensity of any 84.5~GHz emission is lower relative to the 36.2~GHz emission when compared to MM4.

Maser models show that both the 36.2 and 84.5~GHz maser transitions respond similarly to the conditions of their environments \citep{McEwen+14, Leurini+16}. Figure 2 of \citet{McEwen+14} shows that across the range of possible environmental densities for these transitions, at the lowest viable densities the 36.2/84.5 line ratio is at a maximum, with this ratio dropping and tending towards zero as density increases. This suggests that in the region we are seeing both of these transitions, the density is likely higher than in the regions of methanol we do not detected the 84.5~GHz transition. This is further supported by the fact that the range of viable densities for the 44.1~GHz maser transition are shifted towards lower values \citep{McEwen+14, Leurini+16}. Therefore, in a higher density environment we would expect to see higher gain from both the 36.2 and 84.5~GHz transitions (compared to the 44.1~GHz line), which is what we observe towards the 84.5~GHz maser region. From \citet{McEwen+14}, the line ratio of $1.2\pm0.4$ we observed between the 36.2 and 84.5~GHz masers in the same region corresponds to a number density of H$_2$ of approximately $10^6$--$10^7$~cm$^{-3}$, whereas the upper limits of between 1.5 and 4.2 we calculated for the remaining regions correspond to densities of $\sim10^3$--$10^6$~cm$^{-3}$ \citep[Figure 2 of][]{McEwen+14}. Maser beaming can significantly affect line ratios between observed methanol maser transitions, especially when considering methanol lines from different transition series \citep{Sobolev+07,Sobolev&18}. However, as the 36.2 and 84.5~GHz transitions are from the same series \cite[first MMI regime from][]{Sobolev+07}, we expect beaming may have less of an affect on this ratio. \citet{Ellingsen+17b} suggest that the class I methanol masers are predominately from low-gain, relatively diffuse emission regions.  So we would expect these masers to be at significantly lower densities than those characterising cold dense Galactic cores. However, such densities could be achieved within dense giant molecular clouds similar to Sagitarius B2 \citep{Huttemeister+93}. It should be noted that the models from both \citet{McEwen+14} and \citet{Leurini+16} focused on investigation of Galactic environments (primarily high-mass star formation regions), and it may be possible there is a low density regime where the 36.2/84.5~GHz line ratio approaches unity which would be more consistent with diffuse maser emission.

\subsection{Future prospects}

Across the 7 discrete methanol masing regions in NGC~253 \citep[as defined by ][]{Ellingsen+17b}, we find three different combinations of the different observed class~I methanol maser transitions (36.2, 44.1 and 84.5~GHz). We have a region (MM4) with all three class~I lines present, a region with only 36.2 and 44.1~GHz lines observed (MM1), and the remaining only display 36.2~GHz emission. As discussed in Section \ref{sec:compare}, differences in the ratios of integrated line intensities of these maser lines indicates differences in the masing environments. Therefore, such diversity in the distribution of this extremely luminous class~I emission across these regions indicates that the conditions at these locations must vary substantially. This outlines the importance of observing as many class~I methanol lines towards these sources as possible in order to obtain the most complete picture of the methanol maser environments.

The high sensitivity and resolution of ALMA presents an opportunity to search for the 84.5~GHz maser transition at much lower luminosities, or towards sources at much larger distances. When ALMA band 1 becomes operational, ALMA will become the best instrument for searching for 36.2 and 44.1~GHz methanol emission towards extragalactic sources. In addition to searching for lines with current extragalactic detections, ALMA is able to observe the next two higher frequency transitions from the same transition series as the 36.2 and 84.5~GHz maser lines (the 133~GHz $7_{-1}\rightarrow6_0$E line in band 4 and 229~GHz $8_{-1}\rightarrow7_0$E line in band 6). In addition to obtaining more observational data on these extragalactic methanol transitions, it is a good time to expand theoretical studies allowing interpretation of the data. These theoretical models, developed using current data, can be subsequently tested by future observations of any predicted lines.

\section{Conclusion} \label{sec:conc}

We present results from 3\,mm methanol maser search towards NGC~253 and identify a region of 84.5~GHz methanol emission at a location with previously observed 36.2 and 44.1~GHz class~I methanol maser emission. The presence of multiple other class~I maser lines at this location, along with a high isotropic luminosity and narrow linewidth indicate this emission is likely the result of a maser process. This is the second detection of 84.5~GHz class~I methanol towards an external galaxy. 
We find a 36.2/84.5~GHz integrated line intensity ratio of $1.2\pm0.4$ at the location of the 84.5~GHz maser which is consistent with the expected line ratios between these transitions in Galactic sources.
The 84.5~GHz maser transition appears to be a promising candidate for comparison with the 36.2~GHz methanol transition in order to probe the fundamental constants of physics on cosmological scales.
We suggest conducting follow up observations of the 84.5~GHz methanol towards NGC~253 and other extragalactic class~I methanol sources. The increased sensitivity of ALMA will allow us to determine how common the presence of 84.5~GHz methanol is towards these sources, and if detected allow for comparison of the masing environments across these sources.

\acknowledgments
We thank the anonymous referee for useful suggestions which helped to improve this paper. The Australia Telescope Compact Array is part of the Australia Telescope National Facility, which is funded by the Commonwealth of Australia for operation as a National Facility managed by CSIRO. This research has made use of NASA's Astrophysics Data System Abstract Service. This research also utilised APLPY, an open-source plotting package for PYTHON hosted at http://aplpy.github.com. This research made use of Astropy, a community-developed core Python package for Astronomy \citep{astropy+13}.

\end{document}